\documentclass[journal=jacsat,manuscript=article]{achemso}

\usepackage[version=3]{mhchem} 
\usepackage{xcolor}
\usepackage{amsfonts}
\usepackage{braket}
\usepackage{hyperref}

\author{Swapnil Patel}
\affiliation{Duke Quantum Center, Duke University, Durham, NC 27701, USA}
\alsoaffiliation{Department of Physics, Duke University, Durham, NC 27708, USA}

\author{Kenneth R. Brown}
\email{kenneth.r.brown@duke.edu}
\affiliation{Duke Quantum Center, Duke University, Durham, NC 27701, USA}
\alsoaffiliation{Department of Physics, Duke University, Durham, NC 27708, USA}
\alsoaffiliation{Departments of Electrical and Computer Engineering and Chemistry, Duke University, Durham, NC 27708, USA}

\title[An \textsf{achemso} demo]
  {Precise Determination of Excited State Rotational Constants and Black-Body Thermometry in Coulomb Crystals of Ca$^+$ and CaH$^+$}

\abbreviations{IR,NMR,UV}
\keywords{American Chemical Society, \LaTeX}

\begin{document}




\begin{abstract}

    We present high-resolution rovibronic spectroscopy of calcium monohydride molecular ions (CaH\textsuperscript{+}) co-trapped in a Coulomb crystal with calcium ions (\textsuperscript{40}Ca\textsuperscript{+}), focusing on rotational transitions in the $\ket{X^1\Sigma^+, \nu" = 0} \rightarrow \ket{A^1\Sigma^+, \nu' = 2}$ manifold. By resolving individual P and R branch transitions with record precision and using Fortrat analysis, we extract key spectroscopic constants for the excited state $\ket{A^1\Sigma^+, \nu' = 2}$, specifically, the band origin, the rotational constant, and the centrifugal correction. Additionally, we demonstrate the application of high-resolution rotational spectroscopy of CaH\textsuperscript{+} presented here as an in-situ probe of local environmental temperature. We correlate the relative amplitudes of the observed transitions to the underlying thermalized ground-state rotational population distribution and extract the black-body radiation (BBR) temperature.
    
\end{abstract}


\section{Introduction}

The use of ion traps for molecular ions, often with co-trapped atomic ions, provides a controlled and isolated environment for probing the complex internal structure of molecules \cite{calvin2018spectroscopy, willitsch2012coulomb}. Trapped and laser-cooled atomic ions not only facilitate the formation of molecular ions through reactions with relevant gases but also sympathetically cool the translational motion of the co-trapped molecular ions \cite{rugango2015}. This low-temperature, low-background environment, combined with long trapping lifetimes, has allowed for precision spectroscopy and the manipulation of the internal states of molecular ions \cite{calvin2018spectroscopy, willitsch2012coulomb, schiller2022precision}. High-resolution molecular spectroscopy enabled by trapped ions has broad applications, including for time variation of fundamental constants, tests of standard model and beyond standard model theories, identifying molecular species in astrophysical objects, and as tests of advanced quantum chemistry methods \cite{safronova2018search, patra2020proton, alighanbari2020precise, kortunov2021proton, roussy2023improved, schenkel2024laser, demille2024quantum}. 

Among the various methods to explore the structure of molecules in ion traps, a promising development is quantum logic spectroscopy (QLS), which enables probing and coherent manipulation of molecular internal structures in their ground electronic state \cite{schmidt2005spectroscopy, sinhal2023molecular, chou2017preparation}. QLS has been demonstrated with unprecedented precision in experiments involving a Ca\textsuperscript{+}-CaH\textsuperscript{+} ion chain and been adapted to other molecular ions of interest \cite{wolf2016non, sinhal2020quantum, lin2020quantum, chou2020frequency, liu2024quantum, holzapfel2024quantum}. Despite the high resolution achievable with QLS methods, these techniques can be challenging to implement and are best suited for precision spectroscopy and quantum manipulation of molecular ions in their ground state. In contrast, resonant photodissociation spectroscopy is a simpler and more versatile method for probing transitions in molecular ions, as it allows exploration of transitions within and between electronic energy levels. While the technique's destructive nature is a notable downside, it often facilitates easier experimental implementation \cite{roth2006rovibrational, hojbjerre2009rotational, rugango2016vibronic, calvin2018rovibronic, wu2024photodissociation}.

In this study, we leverage the precise knowledge of the ground state of CaH\textsuperscript{+}, obtained through QLS by the NIST molecular ion group \cite{chou2020frequency}, to perform high-resolution spectroscopy of a rovibronic band connecting the ground and first excited electronic states. By accurately determining the rotational constants of this band, our measurements provide crucial experimental data on excited electronic states that is often lacking, aiding in the testing and validation of ab-initio quantum chemistry methods for molecules in excited states.

Specifically, we report on resonance-enhanced multi-photon dissociation spectroscopy (REMPD) of individual rotational transitions in the $\ket{X^1\Sigma^+, \nu" = 0} \rightarrow \ket{A^1\Sigma^+, \nu' = 2}$ band. We resolve P and R branch transitions to better than 0.1 cm\textsuperscript{-1} for high J states and better than 0.3 cm\textsuperscript{-1} for low J states, allowing us to assign ground and excited state rotational quantum numbers and fit a Fortrat parabola to extract the band head, the excited state rotational constant, and the centrifugal correction with high precision. Additionally, we use this technique to perform black-body radiation (BBR) thermometry within our setup, determining the local environmental temperature by probing the rotational state distribution in the ground electronic ground vibrational state resulting from the coupling of the permanent dipole moment of the molecule with the BBR field \cite{koelemeij2007blackbody}. Information about the local temperature is crucial to lower systematic uncertainties in atomic clocks and experiments involving molecules precisely due to the thermalization of internal states which affects state preparation and measurement fidelities \cite{bertelsen2006rotational, koelemeij2007blackbody, vanhaecke2007precision, beloy2014atomic, norrgard2021quantum, liu2024quantum}. This method of in-situ BBR thermometry via rotational transitions in CaH\textsuperscript{+} offers a straightforward and effective approach for probing the BBR field.


\section{Methods}

A 3D crystal of approximately 130 calcium ions (\textsuperscript{40}Ca\textsuperscript{+}) is trapped in a linear Paul trap with an RF drive frequency $\mathrm{\Omega = 1.7 \, MHz}$ and an $\mathrm{r_0 = 9.6 \, mm}$ \cite{jyothi2019}. The calcium ions are Doppler cooled using 397 nm cooling and 866 nm repump lasers. Calcium monohydride molecular ions (\textsuperscript{40}CaH\textsuperscript{+}) are formed by the reaction of the trapped Ca\textsuperscript{+} ions with a gas of hydrogen (H\textsubscript{2}) introduced into the chamber through a real-time, rate controlled piezo-electric leak valve (Oxford Applied Research PLV-1000). H\textsubscript{2} gas is leaked at a pressure of approximately $ \mathrm{4.5 \times 10^{-8} \; torr}$ (base pressure: $\mathrm{9.0 \times 10^{-10} \; torr}$) for 4 minutes with the aim of converting about 25\% of the calcium crystal to CaH\textsuperscript{+}. The translational motion of the molecular ions is sympathetically cooled by the co-trapped, laser-cooled calcium ions \cite{rugango2015}. We wait for a few seconds before switching on the dissociation laser to allow the internal states of the molecules to thermalize due to black-body radiation in our room temperature system. The experiment control for work presented here is written using Duke ARTIQ Extensions (DAX), an open-source, modular, real-time quantum control software framework built on top of ARTIQ (M-Labs) \cite{RiesebosIEEE22, riesebos2021duke, RiesebosIEEE22_1}.

A frequency-doubled continuous-wave (CW) Titanium-Sapphire laser (MSquared SolsTiS ECD-X), with the LBO crystal cut for a center wavelength of 380 nm, is used for multi-photon dissociation spectroscopy of CaH\textsuperscript{+}. The tilt of the LBO crystal within the frequency doubling cavity allows for wavelength tunability of $\mathrm{~ \pm 5  \, nm}$ which allows us to target the second vibrational level within the first excited electronic state ($\ket{A^1\Sigma^+, \nu' = 2}$) of CaH\textsuperscript{+}. The laser linewidth was measured to be approximately 50 kHz for a fundamental Ti-Sapphire wavelength of 729 nm using a transfer cavity normally used for addressing the \textsuperscript{40}Ca\textsuperscript{+} \textsuperscript{2}S\textsubscript{1/2}-\textsuperscript{2}D\textsubscript{5/2} quadrupole transition. The dissociation beam is aligned along the axial direction of the trap, with a beam intensity of $\mathrm{2.38 \times 10^4 \, W/m^2}$ and a beam diameter of $\mathrm{463 \, \mu m}$ to address the whole 3D ion crystal. REMPD operates by first resonantly exciting the \textsuperscript{40}CaH\textsuperscript{+} molecular ions on a specific rovibronic transition - in this case, the $\ket{X^1\Sigma^+, \nu" = 0} \rightarrow \ket{A^1\Sigma^+, \nu' = 2}$ manifold - while a second photon of the same wavelength excites the molecular ion to a dissociative level which asymptotes to \textsuperscript{40}Ca\textsuperscript{+}(\textsuperscript{2}P) + H(\textsuperscript{1}S) \cite{abe2012ab}. The dissociated \textsuperscript{40}Ca\textsuperscript{+} ion is re-trapped and cooled, and the recovered fluorescence is a direct measure of the detuning from the resonance of a specific rotational transition.

\section{Results and discussion}

\begin{figure}[h!]
    \includegraphics[width=0.97\linewidth]{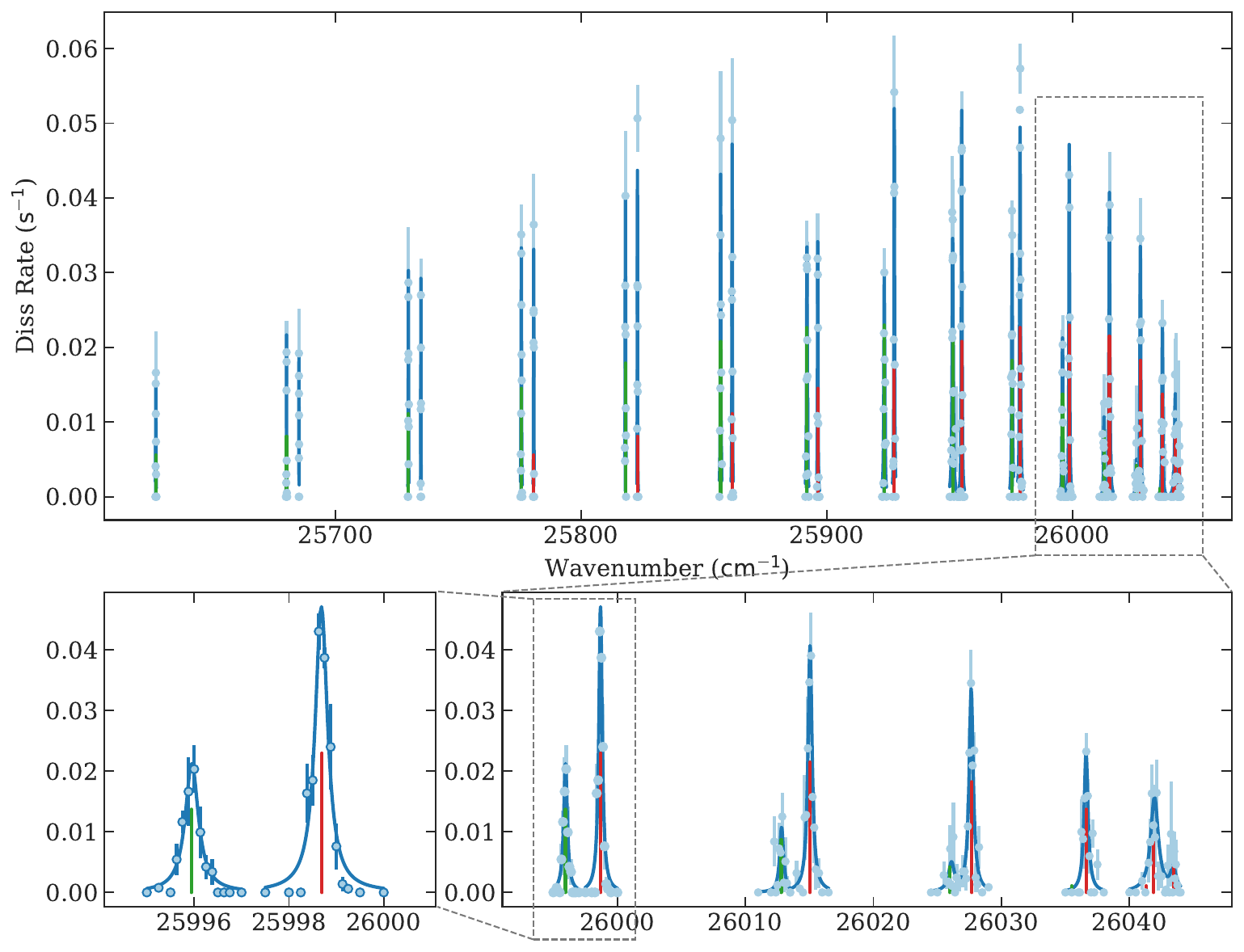}
    \caption{Rotational spectroscopy of CaH\textsuperscript{+} using REMPD. Light blue points represent the mean and one standard error from five samples, fit to a Lorentzian to extract the resonant frequency and amplitude (dark blue line). Green/red lines indicate the predicted P/R transitions based on constants from a global fit to Equation \ref{eqn:fortrat}. The line amplitudes reflect the fractional population of the ground rotational state J for each branch, globally scaled to the observed dissociation rate.}
  \label{fig:specdata}
\end{figure}

Our approach is guided by previous studies on vibronic and rovibronic spectroscopy of CaH\textsuperscript{+} using REMPD \cite{rugango2016vibronic, calvin2018rovibronic}. However, the broad linewidth of the pulsed laser in those experiments, along with discrepancies between the experimental data and predicted transitions, limits the reliability of these frequencies beyond serving as a starting point. Indeed, we eventually find a discrepancy of approximately 80 cm\textsuperscript{-1} from the previous data and attribute it to the imprecise nature of the spectrometer-frequency correlation used in earlier experiments \cite{calvin2018rovibronic}. 

The dissociation laser is scanned over a range of 383–390 nm, limited by the range of the LBO crystal in the frequency-doubling cavity. An initial coarse scan is performed at high power with a resolution of 0.8 cm\textsuperscript{-1} to identify regions of interest. Subsequently, a fine scan is conducted in these regions, and dissociation rates are extracted by monitoring the recovery of the calcium ion fluorescence. For a given dissociation laser frequency, the background-subtracted calcium ion fluorescence as a function of dissociation time is fit to a first-order rate equation, $\mathrm{A(t) = A (1 - e^{-\Gamma(\nu) t})}$, where A(t) is the fluorescence at time t, A is the steady state fluorescence, and the rate $\Gamma(\nu)$ depends on the resonance frequency of the transitions and the population distribution of the ground rotational states.

The dissociation rates, $\Gamma(\nu)$, as a function of dissociation frequency, $\nu$, are fit to a Lorentzian to extract each transition's resonance frequency, amplitude, and width. The linewidth of our laser ($\mathrm{1.7 \times 10^{-4} \, cm^{-1}}$) and the predicted linewidth of the resonant transition ($\mathrm{7 \times 10^{-2} \, cm^{-1}}$) are both narrower than the observed linewidth of the transition. Therefore, we attribute the broader observed linewidth to power broadening and to the second stage process of REMPD. The narrow linewidth achieved through the CW laser, as evidenced by the width of the Lorentzians fit to each transition (Figure \ref{fig:specdata}), is instrumental in resolving the individual P and R branches. Shown in Figure \ref{fig:specdata} is the complete spectrum of the $\ket{X^1\Sigma^+, \nu" = 0, J"} \rightarrow \ket{A^1\Sigma^+, \nu' = 2, J'}$ manifold that is addressable, including transitions near the band head and a specific set of P and R transitions to illustrate the experimental resolution achieved.

\subsection{Determination of excited state rotational constants}

The experimental transition frequencies are initially assigned to a set of ground rotational states by considering the intensity of the observed peaks relative to the thermalized population distribution of the ground states. Since we are dealing with a $^1\Sigma \rightarrow$$^1\Sigma$ transition, only the P ($\Delta J = -1$) and R ($\Delta J = +1$) branch transitions are allowed, which can be described by the following single equation. 
\begin{equation}
    \mathrm{\nu_{obs} = \nu_0 + (B' + B") \,m + (B' - B" - D' + D")\,m^2 - 2\,(D' +D")\,m^3 - (D' - D")\,m^4} 
    \label{eqn:fortrat}
\end{equation}

where m is an integer representing the branch and the rotational quantum number information. Specifically, $\mathrm{m = -J}$ for the P branch and $\mathrm{m = J + 1}$ for the R branch, where J is the ground state rotational quantum number. $\nu_0$ represents the transition frequency between the vibronic levels (the band origin), while B' and D' (B" and D") are the rotational constants and the centrifugal correction terms for the excited (ground) vibronic state.

\begin{figure}[h]
    \includegraphics[width=0.7\linewidth]{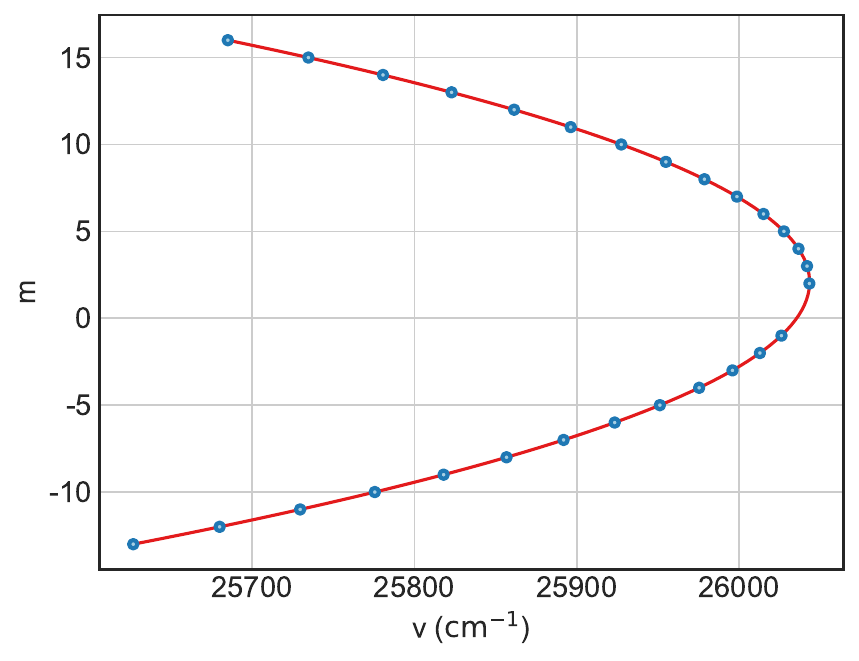}
  \caption{Fortrat diagram for the $\ket{X^1\Sigma^+, \nu" = 0, J"} \rightarrow \ket{A^1\Sigma^+, \nu' = 2, J'}$ manifold. Resonant transition frequencies determined from Lorentzian fits are shown in blue, with error bars (1 standard deviation of fit error) too small to be visible. A global fit to the data, based on Equation \ref{eqn:fortrat}, is shown in red. Resulting fit parameters are provided in Table \ref{tbl:constants}.}
  \label{fig:fortratfit}
\end{figure}

Initially, observed transitions near the band-head and an initial J assignment are used to fit a Fortrat curve (Equation \ref{eqn:fortrat}). The J-state assignments for the observed resonant transition frequencies are then permuted to minimize the errors in the global fit. The assignment with the lowest error is considered the true J-state configuration. This provides us with a prediction of all the remaining transition frequencies. Finally, all the observed transitions, along with their assignments, are fit to Equation \ref{eqn:fortrat} to extract the rotational constants (Figure \ref{fig:fortratfit}). The fit incorporates the precisely measured ground state constants from NIST - $\mathrm{B" = 4.753347661 (57) \,cm^{-1}}$ and $\mathrm{D" = 1.938731 (63) \times 10^{-4} \,cm^{-1}}$ \cite{chou2020frequency}. The narrow, resolved transitions significantly constrain the global fit, allowing us to determine the rotational constants for the excited state $\ket{A^1\Sigma^+, \nu' = 2}$ with high precision (Table \ref{tbl:constants}).

\begin{table}
  \caption{Experimentally determined constants for the $\ket{A^1\Sigma^+, \nu' = 2}$ excited state, obtained from a global fit to the observed transition frequencies in the $\ket{X^1\Sigma^+, \nu" = 0}$ $ \rightarrow$ $ \ket{A^1\Sigma^+, \nu' = 2}$ rovibrational manifold using Equation \ref{eqn:fortrat}.}
  \label{tbl:constants}
  \begin{tabular}{ll}
    \hline
    Band Origin* ($\nu_0$) & 26035.4887(72)   \\
    Rotational Constant (B)  & 2.91150(10)  \\
    Centrifugal Correction (D)  & $1.7186(30) \times 10^{-4}$  \\
    \hline
    *$\ket{X^1\Sigma^+, \nu" = 0, J" = 0} \rightarrow \ket{A^1\Sigma^+, \nu' = 2, J' = 0}$ \\
    All values reported in cm\textsuperscript{-1}.
  \end{tabular}
\end{table}
\subsection{Black-body thermometry}

\begin{figure}[h!]
    \includegraphics[width=0.7\linewidth]{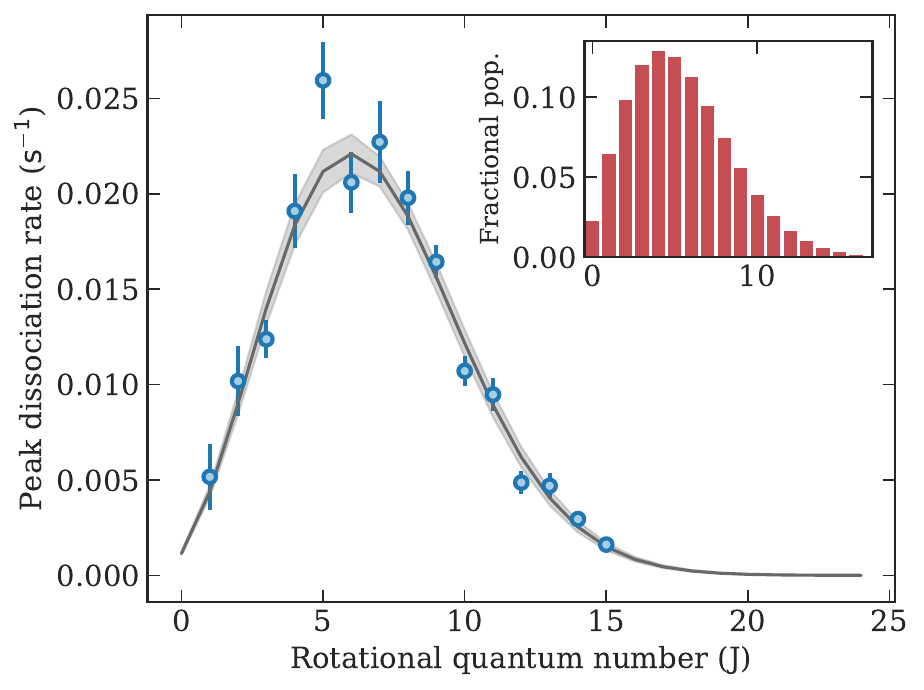}  
  \caption{Peak amplitudes from the Lorentzian fits as a function of the assigned ground state rotational quantum number (blue). The gray line represents the amplitude-scaled fit to Equation \ref{eqn:intensity}, with a best-fit temperature of $\mathrm{308(8)}$ K. Error bars on the data points (blue) and the shaded region around the fit curve (gray) indicate a 1 standard deviation of the fit parameters. Inset: Normalized thermal population distribution of the rotational states in the ground vibrational and ground electronic state of CaH\textsuperscript{+} at 300 K.}
  \label{fig:bbrfit}
\end{figure}

A defining feature of working with molecules is the coupling of the internal degrees of freedom of the molecules to the thermal radiation of the environment. For CaH\textsuperscript{+} at room temperature, this coupling results in populating the first 15 rotational levels to an appreciable amount given by the Boltzmann distribution (Figure \ref{fig:bbrfit} Inset). For molecular ions translationally cooled by co-trapped atomic ions, the Coulomb interaction is too long range to couple to the internal degrees of freedom and thus, the internal population distribution of molecular ions remains themalized to the local environment temperature \cite{bertelsen2006rotational}. As such, the rotation-resolved REMPD spectroscopy established in this work allows us to probe the underlying rotational state distribution in CaH\textsuperscript{+}. From this distribution, we extract the local environment temperature of the ions, based on the interaction of the BBR field with the rotational states of CaH\textsuperscript{+} in its ground electronic ground vibrational level.

To determine the environment temperature, we compare the observed intensity distribution—derived from the amplitude of the Lorentzian fits for each resolved rotational transition—with the expected intensity distribution given by Equation \ref{eqn:intensity}, where the normalization factor A and temperature T are free parameters determined from the fit (Figure \ref{fig:bbrfit}).
\begin{equation}
    \mathrm{I_{J,\nu_0} = A \frac{1}{ \sum_{J}^{} I_J } (J+1)^2 \, exp[-E_J / k_B T]}
    \label{eqn:intensity}
\end{equation}

where $\mathrm{E_J = h [B"J(J+1) - D"J^2(J+1)^2]}$ is the rotational energy of J-th level, B" and D" are ground state rotational constant and centrifugal correction term given in cm\textsuperscript{-1}. 

Equation \ref{eqn:intensity} differs from the Boltzmann distribution of the rotational levels, using the factor $\mathrm{(J+1)^2}$ rather than $\mathrm{(2J+1)}$ to account for the degeneracies and transition dipole moments of the P and R branches \cite{mchale2017molecular}. The temperature determined from the fit, $308(8)$ K, is in reasonable agreement with our room temperature setup. The fit indicates that the observed dissociation rate closely matches the ground-state thermal population, from which we conclude that the second stage of the REMPD process is essentially instantaneous compared to the resonant first stage. This result is in good agreement with previous studies probing the BBR field using molecular ions and confirms that the internal temperature of the molecular ions is unaffected by sympathetic cooling by co-trapped atomic ions \cite{bertelsen2006rotational, koelemeij2007blackbody}. This demonstration highlights the potential of resolved rotational transitions described in this work as an in-situ probe of the local temperature.

\section{Conclusions}
Leveraging the narrow linewidth and wide frequency tuning range of the frequency-doubled CW Ti-Sapphire laser, we achieve the highest resolution to date in the rovibrational spectroscopy of CaH\textsuperscript{+}. We resolve each P and R branch in the $\ket{X^1\Sigma^+, \nu" = 0}$ $ \rightarrow$ $ \ket{A^1\Sigma^+, \nu' = 2}$ manifold up to J = 15, limited by the low thermal population in higher J states and the range of the LBO crystal. This high-resolution spectrum enables us to determine the band origin, the rotational constant, and the centrifugal correction term for the state $\ket{A^1\Sigma^+, \nu' = 2}$ to record precision, supporting tests of excited-state ab-initio methods. This approach can readily be extended to address other vibrational levels within the excited state by using an appropriate LBO crystal in the Ti-Sapphire frequency-doubling cavity.

Additionally, we determine the local environment temperature by examining the coupling of thermal radiation to the rotational population distribution in the ground state of CaH\textsuperscript{+}, establishing CaH\textsuperscript{+} and REMPD as tools to probe in-situ system temperature. Using this technique, we determine the environment temperature that is in reasonable agreement with the expected room temperature of our system. This capability offers a pathway to investigate the effects of ion trap parameters on local temperature \cite{liu2024quantum}, reduce black-body radiation (BBR) related systematics in high-precision clocks \cite{beloy2014atomic}, and study the efficiency of various schemes for cooling the internal states of molecular ions including sympathetic cooling of molecular ions with ultracold trapped neutral atoms\cite{koelemeij2007blackbody, staanum2010rotational, schneider2010all, rellergert2013evidence, lien2014broadband, hansen2014efficient, hudson2009method}. The precision spectroscopy and thermometry method demonstrated here with CaH\textsuperscript{+} can be adapted to any molecular ion system in a Coulomb crystal, either by monitoring the atomic ion fluorescence as shown in this work or, more generally, by observing changes in the crystal structure upon molecular ion dissociation.

\pagebreak

\begin{acknowledgement}

This work was supported by the Army Research Office (W911NF-21-1-0346). The authors thank S. Jyothi for building the experimental setup and Kevin Li for designing the detection system. We also thank Leon Riesebos and Aniket Dalvi for contributing to the experiment control codebase, and Lu Qi and Evan Reed for helpful discussions.

\end{acknowledgement}





\bibliography{achemso-demo}

\providecommand{\latin}[1]{#1}
\makeatletter
\providecommand{\doi}
  {\begingroup\let\do\@makeother\dospecials
  \catcode`\{=1 \catcode`\}=2 \doi@aux}
\providecommand{\doi@aux}[1]{\endgroup\texttt{#1}}
\makeatother
\providecommand*\mcitethebibliography{\thebibliography}
\csname @ifundefined\endcsname{endmcitethebibliography}  {\let\endmcitethebibliography\endthebibliography}{}
\begin{mcitethebibliography}{43}
\providecommand*\natexlab[1]{#1}
\providecommand*\mciteSetBstSublistMode[1]{}
\providecommand*\mciteSetBstMaxWidthForm[2]{}
\providecommand*\mciteBstWouldAddEndPuncttrue
  {\def\EndOfBibitem{\unskip.}}
\providecommand*\mciteBstWouldAddEndPunctfalse
  {\let\EndOfBibitem\relax}
\providecommand*\mciteSetBstMidEndSepPunct[3]{}
\providecommand*\mciteSetBstSublistLabelBeginEnd[3]{}
\providecommand*\EndOfBibitem{}
\mciteSetBstSublistMode{f}
\mciteSetBstMaxWidthForm{subitem}{(\alph{mcitesubitemcount})}
\mciteSetBstSublistLabelBeginEnd
  {\mcitemaxwidthsubitemform\space}
  {\relax}
  {\relax}

\bibitem[Calvin and Brown(2018)Calvin, and Brown]{calvin2018spectroscopy}
Calvin,~A.~T.; Brown,~K.~R. Spectroscopy of molecular ions in Coulomb crystals. \emph{J. Phys. Chem. Lett.} \textbf{2018}, \emph{9}, 5797--5804\relax
\mciteBstWouldAddEndPuncttrue
\mciteSetBstMidEndSepPunct{\mcitedefaultmidpunct}
{\mcitedefaultendpunct}{\mcitedefaultseppunct}\relax
\EndOfBibitem
\bibitem[Willitsch(2012)]{willitsch2012coulomb}
Willitsch,~S. Coulomb-crystallised molecular ions in traps: methods, applications, prospects. \emph{Int. Rev. Phys. Chem.} \textbf{2012}, \emph{31}, 175--199\relax
\mciteBstWouldAddEndPuncttrue
\mciteSetBstMidEndSepPunct{\mcitedefaultmidpunct}
{\mcitedefaultendpunct}{\mcitedefaultseppunct}\relax
\EndOfBibitem
\bibitem[Rugango \latin{et~al.}(2015)Rugango, Goeders, Dixon, Gray, Khanyile, Shu, Clark, and Brown]{rugango2015}
Rugango,~R.; Goeders,~J.~E.; Dixon,~T.~H.; Gray,~J.~M.; Khanyile,~N.; Shu,~G.; Clark,~R.~J.; Brown,~K.~R. Sympathetic cooling of molecular ion motion to the ground state. \emph{New J. Phys.} \textbf{2015}, \emph{17}, 035009\relax
\mciteBstWouldAddEndPuncttrue
\mciteSetBstMidEndSepPunct{\mcitedefaultmidpunct}
{\mcitedefaultendpunct}{\mcitedefaultseppunct}\relax
\EndOfBibitem
\bibitem[Schiller(2022)]{schiller2022precision}
Schiller,~S. Precision spectroscopy of molecular hydrogen ions: an introduction. \emph{Cont. Phys.} \textbf{2022}, \emph{63}, 247--279\relax
\mciteBstWouldAddEndPuncttrue
\mciteSetBstMidEndSepPunct{\mcitedefaultmidpunct}
{\mcitedefaultendpunct}{\mcitedefaultseppunct}\relax
\EndOfBibitem
\bibitem[Safronova \latin{et~al.}(2018)Safronova, Budker, DeMille, Kimball, Derevianko, and Clark]{safronova2018search}
Safronova,~M.; Budker,~D.; DeMille,~D.; Kimball,~D. F.~J.; Derevianko,~A.; Clark,~C.~W. Search for new physics with atoms and molecules. \emph{Rev. Mod. Phys.} \textbf{2018}, \emph{90}, 025008\relax
\mciteBstWouldAddEndPuncttrue
\mciteSetBstMidEndSepPunct{\mcitedefaultmidpunct}
{\mcitedefaultendpunct}{\mcitedefaultseppunct}\relax
\EndOfBibitem
\bibitem[Patra \latin{et~al.}(2020)Patra, Germann, Karr, Haidar, Hilico, Korobov, Cozijn, Eikema, Ubachs, and Koelemeij]{patra2020proton}
Patra,~S.; Germann,~M.; Karr,~J.-P.; Haidar,~M.; Hilico,~L.; Korobov,~V.; Cozijn,~F.; Eikema,~K.; Ubachs,~W.; Koelemeij,~J. Proton-electron mass ratio from laser spectroscopy of HD$^+$ at the part-per-trillion level. \emph{Science} \textbf{2020}, \emph{369}, 1238--1241\relax
\mciteBstWouldAddEndPuncttrue
\mciteSetBstMidEndSepPunct{\mcitedefaultmidpunct}
{\mcitedefaultendpunct}{\mcitedefaultseppunct}\relax
\EndOfBibitem
\bibitem[Alighanbari \latin{et~al.}(2020)Alighanbari, Giri, Constantin, Korobov, and Schiller]{alighanbari2020precise}
Alighanbari,~S.; Giri,~G.; Constantin,~F.~L.; Korobov,~V.; Schiller,~S. \emph{Nature} \textbf{2020}, \emph{581}, 152--158\relax
\mciteBstWouldAddEndPuncttrue
\mciteSetBstMidEndSepPunct{\mcitedefaultmidpunct}
{\mcitedefaultendpunct}{\mcitedefaultseppunct}\relax
\EndOfBibitem
\bibitem[Kortunov \latin{et~al.}(2021)Kortunov, Alighanbari, Hansen, Giri, Korobov, and Schiller]{kortunov2021proton}
Kortunov,~I.; Alighanbari,~S.; Hansen,~M.; Giri,~G.; Korobov,~V.; Schiller,~S. Proton--electron mass ratio by high-resolution optical spectroscopy of ion ensembles in the resolved-carrier regime. \emph{Nat. Phys.} \textbf{2021}, \emph{17}, 569--573\relax
\mciteBstWouldAddEndPuncttrue
\mciteSetBstMidEndSepPunct{\mcitedefaultmidpunct}
{\mcitedefaultendpunct}{\mcitedefaultseppunct}\relax
\EndOfBibitem
\bibitem[Roussy \latin{et~al.}(2023)Roussy, Caldwell, Wright, Cairncross, Shagam, Ng, Schlossberger, Park, Wang, Ye, and Cornell]{roussy2023improved}
Roussy,~T.~S.; Caldwell,~L.; Wright,~T.; Cairncross,~W.~B.; Shagam,~Y.; Ng,~K.~B.; Schlossberger,~N.; Park,~S.~Y.; Wang,~A.; Ye,~J.; Cornell,~E.~A. An improved bound on the electron’s electric dipole moment. \emph{Science} \textbf{2023}, \emph{381}, 46--50\relax
\mciteBstWouldAddEndPuncttrue
\mciteSetBstMidEndSepPunct{\mcitedefaultmidpunct}
{\mcitedefaultendpunct}{\mcitedefaultseppunct}\relax
\EndOfBibitem
\bibitem[Schenkel \latin{et~al.}(2024)Schenkel, Alighanbari, and Schiller]{schenkel2024laser}
Schenkel,~M.; Alighanbari,~S.; Schiller,~S. Laser spectroscopy of a rovibrational transition in the molecular hydrogen ion $\mathrm{N_2 ^+}$. \emph{Nat. Phys.} \textbf{2024}, \emph{20}, 383--388\relax
\mciteBstWouldAddEndPuncttrue
\mciteSetBstMidEndSepPunct{\mcitedefaultmidpunct}
{\mcitedefaultendpunct}{\mcitedefaultseppunct}\relax
\EndOfBibitem
\bibitem[DeMille \latin{et~al.}(2024)DeMille, Hutzler, Rey, and Zelevinsky]{demille2024quantum}
DeMille,~D.; Hutzler,~N.~R.; Rey,~A.~M.; Zelevinsky,~T. Quantum sensing and metrology for fundamental physics with molecules. \emph{Nat. Phys.} \textbf{2024}, 1--9\relax
\mciteBstWouldAddEndPuncttrue
\mciteSetBstMidEndSepPunct{\mcitedefaultmidpunct}
{\mcitedefaultendpunct}{\mcitedefaultseppunct}\relax
\EndOfBibitem
\bibitem[Schmidt \latin{et~al.}(2005)Schmidt, Rosenband, Langer, Itano, Bergquist, and Wineland]{schmidt2005spectroscopy}
Schmidt,~P.~O.; Rosenband,~T.; Langer,~C.; Itano,~W.~M.; Bergquist,~J.~C.; Wineland,~D.~J. Spectroscopy using quantum logic. \emph{Science} \textbf{2005}, \emph{309}, 749--752\relax
\mciteBstWouldAddEndPuncttrue
\mciteSetBstMidEndSepPunct{\mcitedefaultmidpunct}
{\mcitedefaultendpunct}{\mcitedefaultseppunct}\relax
\EndOfBibitem
\bibitem[Sinhal and Willitsch(2023)Sinhal, and Willitsch]{sinhal2023molecular}
Sinhal,~M.; Willitsch,~S. Molecular-Ion Quantum Technologies. \emph{Photonic Quantum Technologies: Science and Applications} \textbf{2023}, \emph{1}, 305--332\relax
\mciteBstWouldAddEndPuncttrue
\mciteSetBstMidEndSepPunct{\mcitedefaultmidpunct}
{\mcitedefaultendpunct}{\mcitedefaultseppunct}\relax
\EndOfBibitem
\bibitem[Chou \latin{et~al.}(2017)Chou, Kurz, Hume, Plessow, Leibrandt, and Leibfried]{chou2017preparation}
Chou,~C.-w.; Kurz,~C.; Hume,~D.~B.; Plessow,~P.~N.; Leibrandt,~D.~R.; Leibfried,~D. Preparation and coherent manipulation of pure quantum states of a single molecular ion. \emph{Nature} \textbf{2017}, \emph{545}, 203--207\relax
\mciteBstWouldAddEndPuncttrue
\mciteSetBstMidEndSepPunct{\mcitedefaultmidpunct}
{\mcitedefaultendpunct}{\mcitedefaultseppunct}\relax
\EndOfBibitem
\bibitem[Wolf \latin{et~al.}(2016)Wolf, Wan, Heip, Gebert, Shi, and Schmidt]{wolf2016non}
Wolf,~F.; Wan,~Y.; Heip,~J.~C.; Gebert,~F.; Shi,~C.; Schmidt,~P.~O. Non-destructive state detection for quantum logic spectroscopy of molecular ions. \emph{Nature} \textbf{2016}, \emph{530}, 457--460\relax
\mciteBstWouldAddEndPuncttrue
\mciteSetBstMidEndSepPunct{\mcitedefaultmidpunct}
{\mcitedefaultendpunct}{\mcitedefaultseppunct}\relax
\EndOfBibitem
\bibitem[Sinhal \latin{et~al.}(2020)Sinhal, Meir, Najafian, Hegi, and Willitsch]{sinhal2020quantum}
Sinhal,~M.; Meir,~Z.; Najafian,~K.; Hegi,~G.; Willitsch,~S. Quantum-nondemolition state detection and spectroscopy of single trapped molecules. \emph{Science} \textbf{2020}, \emph{367}, 1213--1218\relax
\mciteBstWouldAddEndPuncttrue
\mciteSetBstMidEndSepPunct{\mcitedefaultmidpunct}
{\mcitedefaultendpunct}{\mcitedefaultseppunct}\relax
\EndOfBibitem
\bibitem[Lin \latin{et~al.}(2020)Lin, Leibrandt, Leibfried, and Chou]{lin2020quantum}
Lin,~Y.; Leibrandt,~D.~R.; Leibfried,~D.; Chou,~C.-w. Quantum entanglement between an atom and a molecule. \emph{Nature} \textbf{2020}, \emph{581}, 273--277\relax
\mciteBstWouldAddEndPuncttrue
\mciteSetBstMidEndSepPunct{\mcitedefaultmidpunct}
{\mcitedefaultendpunct}{\mcitedefaultseppunct}\relax
\EndOfBibitem
\bibitem[Chou \latin{et~al.}(2020)Chou, Collopy, Kurz, Lin, Harding, Plessow, Fortier, Diddams, Leibfried, and Leibrandt]{chou2020frequency}
Chou,~C.-w.; Collopy,~A.~L.; Kurz,~C.; Lin,~Y.; Harding,~M.~E.; Plessow,~P.~N.; Fortier,~T.; Diddams,~S.; Leibfried,~D.; Leibrandt,~D.~R. Frequency-comb spectroscopy on pure quantum states of a single molecular ion. \emph{Science} \textbf{2020}, \emph{367}, 1458--1461\relax
\mciteBstWouldAddEndPuncttrue
\mciteSetBstMidEndSepPunct{\mcitedefaultmidpunct}
{\mcitedefaultendpunct}{\mcitedefaultseppunct}\relax
\EndOfBibitem
\bibitem[Liu \latin{et~al.}(2024)Liu, Schmidt, Liu, Leibrandt, Leibfried, and Chou]{liu2024quantum}
Liu,~Y.; Schmidt,~J.; Liu,~Z.; Leibrandt,~D.~R.; Leibfried,~D.; Chou,~C.-w. Quantum state tracking and control of a single molecular ion in a thermal environment. \emph{Science} \textbf{2024}, \emph{385}, 790--795\relax
\mciteBstWouldAddEndPuncttrue
\mciteSetBstMidEndSepPunct{\mcitedefaultmidpunct}
{\mcitedefaultendpunct}{\mcitedefaultseppunct}\relax
\EndOfBibitem
\bibitem[Holzapfel \latin{et~al.}(2024)Holzapfel, Schmid, Schwegler, Stadler, Stadler, Ferk, Home, and Kienzler]{holzapfel2024quantum}
Holzapfel,~D.; Schmid,~F.; Schwegler,~N.; Stadler,~O.; Stadler,~M.; Ferk,~A.; Home,~J.~P.; Kienzler,~D. Quantum control of a single $\mathrm{H_2 ^+}$ molecular ion. \emph{arXiv:2409.06495} \textbf{2024}, \relax
\mciteBstWouldAddEndPunctfalse
\mciteSetBstMidEndSepPunct{\mcitedefaultmidpunct}
{}{\mcitedefaultseppunct}\relax
\EndOfBibitem
\bibitem[Roth \latin{et~al.}(2006)Roth, Koelemeij, Daerr, and Schiller]{roth2006rovibrational}
Roth,~B.; Koelemeij,~J.; Daerr,~H.; Schiller,~S. Rovibrational spectroscopy of trapped molecular hydrogen ions at millikelvin temperatures. \emph{Phys. Rev. A} \textbf{2006}, \emph{74}, 040501\relax
\mciteBstWouldAddEndPuncttrue
\mciteSetBstMidEndSepPunct{\mcitedefaultmidpunct}
{\mcitedefaultendpunct}{\mcitedefaultseppunct}\relax
\EndOfBibitem
\bibitem[H{\o}jbjerre \latin{et~al.}(2009)H{\o}jbjerre, Hansen, Skyt, Staanum, and Drewsen]{hojbjerre2009rotational}
H{\o}jbjerre,~K.; Hansen,~A.~K.; Skyt,~P.~S.; Staanum,~P.; Drewsen,~M. Rotational state resolved photodissociation spectroscopy of translationally and vibrationally cold MgH$^+$ ions: toward rotational cooling of molecular ions. \emph{New J. Phys.} \textbf{2009}, \emph{11}, 055026\relax
\mciteBstWouldAddEndPuncttrue
\mciteSetBstMidEndSepPunct{\mcitedefaultmidpunct}
{\mcitedefaultendpunct}{\mcitedefaultseppunct}\relax
\EndOfBibitem
\bibitem[Rugango \latin{et~al.}(2016)Rugango, Calvin, Janardan, Shu, and Brown]{rugango2016vibronic}
Rugango,~R.; Calvin,~A.~T.; Janardan,~S.; Shu,~G.; Brown,~K.~R. Vibronic spectroscopy of sympathetically cooled CaH$^+$. \emph{ChemPhysChem} \textbf{2016}, \emph{17}, 3764--3768\relax
\mciteBstWouldAddEndPuncttrue
\mciteSetBstMidEndSepPunct{\mcitedefaultmidpunct}
{\mcitedefaultendpunct}{\mcitedefaultseppunct}\relax
\EndOfBibitem
\bibitem[Calvin \latin{et~al.}(2018)Calvin, Janardan, Condoluci, Rugango, Pretzsch, Shu, and Brown]{calvin2018rovibronic}
Calvin,~A.~T.; Janardan,~S.; Condoluci,~J.; Rugango,~R.; Pretzsch,~E.; Shu,~G.; Brown,~K.~R. Rovibronic spectroscopy of sympathetically cooled $^{40}$CaH$^+$. \emph{J. Phys. Chem. A} \textbf{2018}, \emph{122}, 3177--3181\relax
\mciteBstWouldAddEndPuncttrue
\mciteSetBstMidEndSepPunct{\mcitedefaultmidpunct}
{\mcitedefaultendpunct}{\mcitedefaultseppunct}\relax
\EndOfBibitem
\bibitem[Wu \latin{et~al.}(2024)Wu, Walser, Podlesnic, Isaza-Monsalve, Mattivi, Mu, Nardi, Gniewek, Tomza, Furey, and Schindler]{wu2024photodissociation}
Wu,~Z.; Walser,~S.; Podlesnic,~V.; Isaza-Monsalve,~M.; Mattivi,~E.; Mu,~G.; Nardi,~R.; Gniewek,~P.; Tomza,~M.; Furey,~B.~J.; Schindler,~P. Photodissociation spectra of single trapped CaOH$^+$ molecular ions. \emph{J. Chem. Phys.} \textbf{2024}, \emph{161}\relax
\mciteBstWouldAddEndPuncttrue
\mciteSetBstMidEndSepPunct{\mcitedefaultmidpunct}
{\mcitedefaultendpunct}{\mcitedefaultseppunct}\relax
\EndOfBibitem
\bibitem[Koelemeij \latin{et~al.}(2007)Koelemeij, Roth, and Schiller]{koelemeij2007blackbody}
Koelemeij,~J.; Roth,~B.; Schiller,~S. Blackbody thermometry with cold molecular ions and application to ion-based frequency standards. \emph{Phys. Rev. A} \textbf{2007}, \emph{76}, 023413\relax
\mciteBstWouldAddEndPuncttrue
\mciteSetBstMidEndSepPunct{\mcitedefaultmidpunct}
{\mcitedefaultendpunct}{\mcitedefaultseppunct}\relax
\EndOfBibitem
\bibitem[Bertelsen \latin{et~al.}(2006)Bertelsen, J{\o}rgensen, and Drewsen]{bertelsen2006rotational}
Bertelsen,~A.; J{\o}rgensen,~S.; Drewsen,~M. The rotational temperature of polar molecular ions in Coulomb crystals. \emph{J. Phys. B} \textbf{2006}, \emph{39}, L83\relax
\mciteBstWouldAddEndPuncttrue
\mciteSetBstMidEndSepPunct{\mcitedefaultmidpunct}
{\mcitedefaultendpunct}{\mcitedefaultseppunct}\relax
\EndOfBibitem
\bibitem[Vanhaecke and Dulieu(2007)Vanhaecke, and Dulieu]{vanhaecke2007precision}
Vanhaecke,~N.; Dulieu,~O. Precision measurements with polar molecules: the role of the black body radiation. \emph{Mol. Phys.} \textbf{2007}, \emph{105}, 1723--1731\relax
\mciteBstWouldAddEndPuncttrue
\mciteSetBstMidEndSepPunct{\mcitedefaultmidpunct}
{\mcitedefaultendpunct}{\mcitedefaultseppunct}\relax
\EndOfBibitem
\bibitem[Beloy \latin{et~al.}(2014)Beloy, Hinkley, Phillips, Sherman, Schioppo, Lehman, Feldman, Hanssen, Oates, and Ludlow]{beloy2014atomic}
Beloy,~K.; Hinkley,~N.; Phillips,~N.~B.; Sherman,~J.~A.; Schioppo,~M.; Lehman,~J.; Feldman,~A.; Hanssen,~L.~M.; Oates,~C.~W.; Ludlow,~A.~D. Atomic clock with $1\times 10^{-18}$ room-temperature blackbody Stark uncertainty. \emph{Phys. Rev. Lett.} \textbf{2014}, \emph{113}, 260801\relax
\mciteBstWouldAddEndPuncttrue
\mciteSetBstMidEndSepPunct{\mcitedefaultmidpunct}
{\mcitedefaultendpunct}{\mcitedefaultseppunct}\relax
\EndOfBibitem
\bibitem[Norrgard \latin{et~al.}(2021)Norrgard, Eckel, Holloway, and Shirley]{norrgard2021quantum}
Norrgard,~E.~B.; Eckel,~S.~P.; Holloway,~C.~L.; Shirley,~E.~L. Quantum blackbody thermometry. \emph{New J. Phys.} \textbf{2021}, \emph{23}, 033037\relax
\mciteBstWouldAddEndPuncttrue
\mciteSetBstMidEndSepPunct{\mcitedefaultmidpunct}
{\mcitedefaultendpunct}{\mcitedefaultseppunct}\relax
\EndOfBibitem
\bibitem[Jyothi \latin{et~al.}(2019)Jyothi, Egodapitiya, Bondurant, Jia, Pretzsch, Chiappina, Shu, and Brown]{jyothi2019}
Jyothi,~S.; Egodapitiya,~K.~N.; Bondurant,~B.; Jia,~Z.; Pretzsch,~E.; Chiappina,~P.; Shu,~G.; Brown,~K.~R. A hybrid ion-atom trap with integrated high resolution mass spectrometer. \emph{Rev. Sci. Instrum.} \textbf{2019}, \emph{90}\relax
\mciteBstWouldAddEndPuncttrue
\mciteSetBstMidEndSepPunct{\mcitedefaultmidpunct}
{\mcitedefaultendpunct}{\mcitedefaultseppunct}\relax
\EndOfBibitem
\bibitem[Riesebos \latin{et~al.}(2022)Riesebos, Bondurant, Whitlow, Kim, Kuzyk, Chen, Phiri, Wang, Fang, Horn, Kim, and Brown]{RiesebosIEEE22}
Riesebos,~L.; Bondurant,~B.; Whitlow,~J.; Kim,~J.; Kuzyk,~M.; Chen,~T.; Phiri,~S.; Wang,~Y.; Fang,~C.; Horn,~A.~V.; Kim,~J.; Brown,~K.~R. Modular software for real-time quantum control systems. 2022 IEEE (QCE). 2022; pp 545--555\relax
\mciteBstWouldAddEndPuncttrue
\mciteSetBstMidEndSepPunct{\mcitedefaultmidpunct}
{\mcitedefaultendpunct}{\mcitedefaultseppunct}\relax
\EndOfBibitem
\bibitem[Riesebos \latin{et~al.}(2021)Riesebos, Bondurant, and Brown]{riesebos2021duke}
Riesebos,~L.; Bondurant,~B.; Brown,~K. Duke artiq extensions (dax). \emph{Duke artiq extensions (dax), 2021,[online] Available: https://gitlab.com/duke-artiq/dax} \textbf{2021}, \relax
\mciteBstWouldAddEndPunctfalse
\mciteSetBstMidEndSepPunct{\mcitedefaultmidpunct}
{}{\mcitedefaultseppunct}\relax
\EndOfBibitem
\bibitem[Riesebos and Brown(2022)Riesebos, and Brown]{RiesebosIEEE22_1}
Riesebos,~L.; Brown,~K.~R. Functional simulation of real-time quantum control software. 2022 IEEE (QCE). 2022; pp 535--544\relax
\mciteBstWouldAddEndPuncttrue
\mciteSetBstMidEndSepPunct{\mcitedefaultmidpunct}
{\mcitedefaultendpunct}{\mcitedefaultseppunct}\relax
\EndOfBibitem
\bibitem[Abe \latin{et~al.}(2012)Abe, Moriwaki, Hada, and Kajita]{abe2012ab}
Abe,~M.; Moriwaki,~Y.; Hada,~M.; Kajita,~M. Ab initio study on potential energy curves of electronic ground and excited states of $^{40}$CaH$^+$ molecule. \emph{Chem. Phys. Lett.} \textbf{2012}, \emph{521}, 31--35\relax
\mciteBstWouldAddEndPuncttrue
\mciteSetBstMidEndSepPunct{\mcitedefaultmidpunct}
{\mcitedefaultendpunct}{\mcitedefaultseppunct}\relax
\EndOfBibitem
\bibitem[McHale(2017)]{mchale2017molecular}
McHale,~J.~L. \emph{Molecular spectroscopy}; CRC Press, 2017\relax
\mciteBstWouldAddEndPuncttrue
\mciteSetBstMidEndSepPunct{\mcitedefaultmidpunct}
{\mcitedefaultendpunct}{\mcitedefaultseppunct}\relax
\EndOfBibitem
\bibitem[Staanum \latin{et~al.}(2010)Staanum, H{\o}jbjerre, Skyt, Hansen, and Drewsen]{staanum2010rotational}
Staanum,~P.~F.; H{\o}jbjerre,~K.; Skyt,~P.~S.; Hansen,~A.~K.; Drewsen,~M. Rotational laser cooling of vibrationally and translationally cold molecular ions. \emph{Nat. Phys.} \textbf{2010}, \emph{6}, 271--274\relax
\mciteBstWouldAddEndPuncttrue
\mciteSetBstMidEndSepPunct{\mcitedefaultmidpunct}
{\mcitedefaultendpunct}{\mcitedefaultseppunct}\relax
\EndOfBibitem
\bibitem[Schneider \latin{et~al.}(2010)Schneider, Roth, Duncker, Ernsting, and Schiller]{schneider2010all}
Schneider,~T.; Roth,~B.; Duncker,~H.; Ernsting,~I.; Schiller,~S. All-optical preparation of molecular ions in the rovibrational ground state. \emph{Nat. Phys.} \textbf{2010}, \emph{6}, 275--278\relax
\mciteBstWouldAddEndPuncttrue
\mciteSetBstMidEndSepPunct{\mcitedefaultmidpunct}
{\mcitedefaultendpunct}{\mcitedefaultseppunct}\relax
\EndOfBibitem
\bibitem[Rellergert \latin{et~al.}(2013)Rellergert, Sullivan, Schowalter, Kotochigova, Chen, and Hudson]{rellergert2013evidence}
Rellergert,~W.~G.; Sullivan,~S.~T.; Schowalter,~S.~J.; Kotochigova,~S.; Chen,~K.; Hudson,~E.~R. Evidence for sympathetic vibrational cooling of translationally cold molecules. \emph{Nature} \textbf{2013}, \emph{495}, 490--494\relax
\mciteBstWouldAddEndPuncttrue
\mciteSetBstMidEndSepPunct{\mcitedefaultmidpunct}
{\mcitedefaultendpunct}{\mcitedefaultseppunct}\relax
\EndOfBibitem
\bibitem[Lien \latin{et~al.}(2014)Lien, Seck, Lin, Nguyen, Tabor, and Odom]{lien2014broadband}
Lien,~C.-Y.; Seck,~C.~M.; Lin,~Y.-W.; Nguyen,~J.~H.; Tabor,~D.~A.; Odom,~B.~C. Broadband optical cooling of molecular rotors from room temperature to the ground state. \emph{Nat. Commun.} \textbf{2014}, \emph{5}, 4783\relax
\mciteBstWouldAddEndPuncttrue
\mciteSetBstMidEndSepPunct{\mcitedefaultmidpunct}
{\mcitedefaultendpunct}{\mcitedefaultseppunct}\relax
\EndOfBibitem
\bibitem[Hansen \latin{et~al.}(2014)Hansen, Versolato, K{\l}osowski, Kristensen, Gingell, Schwarz, Windberger, Ullrich, L{\'o}pez-Urrutia, and Drewsen]{hansen2014efficient}
Hansen,~A.~K.; Versolato,~O.; K{\l}osowski,~{\L}.; Kristensen,~S.; Gingell,~A.; Schwarz,~M.; Windberger,~A.; Ullrich,~J.; L{\'o}pez-Urrutia,~J.~C.; Drewsen,~M. Efficient rotational cooling of Coulomb-crystallized molecular ions by a helium buffer gas. \emph{Nature} \textbf{2014}, \emph{508}, 76--79\relax
\mciteBstWouldAddEndPuncttrue
\mciteSetBstMidEndSepPunct{\mcitedefaultmidpunct}
{\mcitedefaultendpunct}{\mcitedefaultseppunct}\relax
\EndOfBibitem
\bibitem[Hudson(2009)]{hudson2009method}
Hudson,~E.~R. Method for producing ultracold molecular ions. \emph{Phys. Rev. A} \textbf{2009}, \emph{79}, 032716\relax
\mciteBstWouldAddEndPuncttrue
\mciteSetBstMidEndSepPunct{\mcitedefaultmidpunct}
{\mcitedefaultendpunct}{\mcitedefaultseppunct}\relax
\EndOfBibitem
\end{mcitethebibliography}

\end{document}